# Anderson localization of solitons in optical lattices with random frequency modulation

Yaroslav V. Kartashov[1] and Victor A. Vysloukh[2]

[1]*ICFO-Institut de Ciencies Fotoniques and Universitat Politecnica de Catalunya, 08034, Barcelona, Spain*
[2]*Departamento de Fisica y Matematicas, Universidad de las Americas - Puebla, Santa Catarina Martir, 72820, Puebla, Mexico*

We report on phenomenon of Anderson-type localization of walking solitons in optical lattices with random frequency modulation, manifested as dramatic enhancement of soliton trapping probability on lattice inhomogeneities with growth of the frequency fluctuation level. The localization process is strongly sensitive to the lattice depth since in shallow lattices walking solitons experience random refraction and/or multiple scattering in contrast to relatively deep lattices, where solitons are typically immobilized in the vicinity of local minimums on modulation frequency.

*PACS numbers: 42.65.Tg, 42.65.Jx, 42.65.Wi.*

The concept of Anderson localization was introduced in the field of solid-state physics for the phenomenon of disorder-induced metal-insulator transition in electronic systems. Anderson localization refers to the situation where electron, when released in a random medium, may stay close to the initial point [1]. The mechanism behind this property has been attributed to multiple scattering of electrons by the random potentials, a feature of the wave nature of electrons. The localization concept may be applied to the classical linear wave systems as well [2]. However the challenging problem is the exploration of the nonlinear analogs of Anderson localization. The interplay between disorder and nonlinearity was recently studied in the systems described by one-dimensional nonlinear Schrödinger equation with random-point impurities [3,4], in discrete waveguide arrays [5,6], and in globular protein [7]. Recently, the concept of nonlinear optical lattices has attracted steady attention in both nonlinear optics and matter waves fields. It was demonstrated that periodic lattices are capable to support stable solitons, whose properties can be tuned continuously from that typical for solitons in uniform nonlinear medium to those representative for solitons in discrete systems by changing strength of the optical lattice [8-11]. The tunability of optical lattice proved to be very promising for soliton management, including radiative switching and parametric steering [12-15]. It is worth noticing that mathematically analogous problems appear in the theory of Bose-Einstein condensates [16-18]. However, up to date theoretical studies of soliton propagation in optical lattices were restricted mainly to the case of regular, perfectly periodic lattices.

In this paper we demonstrate the existence of nonlinear analog of Anderson localization for walking solitons in optical lattices with random frequency modulation. We show that with growth of frequency fluctuations the multiple-scattering scenario gradually replaces the random-refraction one and finally probability of soliton trapping grows dramatically. In terms of solid-state physics this might be interpreted as smooth phase transition between *soliton conductor* and *soliton insulator*. Disorder-induced soliton localization depends strongly on the depth of the lattice and can be tuned.

The generic equation describing the evolution of optical (matter) wave packets in the presence of Kerr (mean field cubic) nonlinearity and a periodic potential induced by a weak optical lattice is the nonlinear Schrödinger equation:

$$i\frac{\partial q}{\partial \xi} = -\frac{1}{2}\frac{\partial^2 q}{\partial \eta^2} - q|q|^2 - pR(\eta)q. \tag{1}$$

In the optical context $q$ is the dimensionless amplitude of light field; the longitudinal and transverse coordinates $\xi, \eta$ are scaled to the diffraction length and the input beam width, respectively. The waveguiding parameter $p$ is proportional to the depth of refractive index modulation, while the random function $R(\eta)$ stands for the transverse profile of refractive index. We assume that the depth of refractive index modulation is small compared to the unperturbed index and is of the order of nonlinear contribution to refractive index due to Kerr effect. In the particular case of optical lattice induction in photorefractive SBN crystal (electro-optic coefficient $r = 1.8 \times 10^{-10}$ m/V, linear refractive index $n_0 = 2.33$) biased with dc static electric field $E_0 \sim 10^5$ V/m, for laser beams with width 10 $\mu$m at wavelength $\lambda = 0.63$ $\mu$m the propagation distance $\xi = 1$ corresponds to 0.6 mm of actual crystal length, while dimensionless amplitude $q \sim 1$ corresponds to real peak intensities about 50 mW/cm$^2$. Notice that at high levels of the uniform background illumination nonlinearity saturation intrinsic for the photorefractive medium can be reduced substantially, and under this conditions Eq. (1) can be readily applied to study general soliton properties.

In the matter waves context Eq. (1) describes dynamics of a one-dimensional Bose-Einstein condensate confined in an optical lattice generated by means of a standing laser wave of wavelength $\lambda$. Now $q$ stands for wave-function, variable $\xi$ stands for time in units of $\tau = 2m\lambda^2/\pi h$, with $m$ being the mass of the atoms and $h$ the Planck's constant, $\eta$ is the longitudinal coordinate along the axis of the quasi-one-dimensional condensate expressed in units of $\lambda \pi^{-1}$. For typical experiments $\lambda$ ranges from 0.8 to 3.2 $\mu$m. Parameter $p$ is proportional to lattice depth $E_0$ expressed in units of the recoil energy $E_{\text{rec}} = h^2/2m\lambda^2$. The lattice depths $p \leq 10$ were already achieved experimentally [16].

Eq. (1) admits several conserved quantities, including the total energy flow $U = \int_{-\infty}^{\infty} |q|^2 \, d\eta$.

Further we consider lattices with random frequency modulation (FM) whose profile is described by the function $R(\eta) = \cos(\Omega \eta)$, where $\Omega(\eta) = [1 + \sigma \rho(\eta)]\Omega_0$, $\Omega_0$ is the carrying frequency, $\rho(\eta)$ is a random process with the Gaussian statistics at fixed $\eta$, zero mean value $\langle \rho \rangle = 0$, and unit variance $\langle \rho^2 \rangle = 1$ (angular brackets stand for the statistical averaging). Parameter $\sigma$ defines the depth of random frequency modulation. The correlation function $\langle \rho(\eta_1)\rho(\eta_2) \rangle = \exp[-(\eta_1 - \eta_2)^2/L_{\text{cor}}^2]$ is assumed to be Gaussian with the correlation length $L_{\text{cor}} > 2\pi/\Omega_0$. The model addressed here is relevant both in the context of optically induced lattices, where certain level of frequency fluctuations is inevitable due to imperfections of lattice-forming waves, and in the context of preformed waveguide arrays, where fluctuations appear upon fabrication. Notice that optically induced lattices with fluctuating spatial frequency actually distort upon evolution, but the rate of distortion decreases dramatically with growth of the correlation length.

When tilted soliton $q(\eta, \xi = 0) = \chi \operatorname{sech}(\chi\eta) \exp(i\alpha_0\eta)$ ($\alpha_0$ is the input angle and $\chi$ is the form-factor) is launched into a regular lattice, it propagates across it provided that $\alpha_0 > \alpha_{\mathrm{cr}}$ where the critical angle is given by $\alpha_{\mathrm{cr}} = 2[p(\pi\Omega_0/2\chi)/\sinh(\pi\Omega_0/2\chi)]^{1/2}$ [16]. If the correlation length is large enough (rather slow frequency modulation), then this equation can be applied to FM lattice as well, if one replaces carrying frequency $\Omega_0$ with instantaneous one $\Omega(\eta)$. The critical angle can be considered as a random function of $\eta$ that acquires maximal values in minima of $\Omega(\eta)$. The walking soliton might then be trapped in the vicinity of minimum of the instantaneous frequency (potential hole) or scattered by a potential barrier-type inhomogeneity. When the input angle greatly exceeds the critical one, the instantaneous tilt angle is given by $\alpha(\eta, \xi) \approx \alpha_0(1 - [\alpha_{\mathrm{cr}}^2(\eta)/2\alpha_0^2]\sin^2[\alpha_0\Omega(\eta)\xi/2])$. In this case probability of soliton scattering or trapping is negligible and it follows a slightly perturbed linear trajectory (the random-refraction scenario). Notice also that radiation, which unavoidably appears when soliton crosses lattice channels, can be neglected if the instantaneous propagation angle is far from the Bragg one $\alpha \ll \alpha_{\mathrm{B}} = \Omega_0/2$ and the carrying spatial frequency is high enough $\Omega_0 \gg \chi^{-1}$.

In numerical simulations we used the Monte-Carlo approach and integrated Eq. (1) with the split-step Fourier method up to the distance $L = 10^2$ for different sets of computer generated random realizations of lattice profiles $R_k(\eta)$, $k = 1,...,N$. We calculated the trajectories of the integral soliton center $\eta_k(\xi) = \int_{-\infty}^{\infty} \eta |q(\eta,\xi)|^2 d\eta / U$ as well as path-averaged soliton center displacement $\eta_k^{\mathrm{av}} = \int_{-\infty}^{\infty} \eta_k(\xi) d\xi / L$ and its squared deviation $S_k^{\mathrm{av}} = \int_{-\infty}^{\infty} [\eta_k(\xi) - \eta_k^{\mathrm{av}}]^2 d\xi / L$. Statistical averaging $\langle \eta \rangle$, $\langle S \rangle$ of these parameters (normalized by their values in the regular lattice $\eta_0 = \alpha_0 L/2$ and $S_0 = \alpha_0^2 L^2/12$) provides information about soliton localization. Key variables are the parameter $p$ that tunes lattice properties and the deviation of spatial frequency fluctuations that defines the disorder level. The correlation length was set to $L_{\mathrm{cor}} = 2$ to establish smooth variation of the instantaneous frequency and statistical averaging was carried out over $N = 10^3$ realizations of lattice profiles.

First we analyzed an impact of lattice depth on soliton localization at moderate disorder level $\sigma \sim 0.1$. In shallow randomly modulated lattices with $p \sim 0.1$ soliton follows slightly perturbed linear trajectory since condition $\alpha_0 > \alpha_{\mathrm{cr}}(\eta)$ holds. However, with growth of lattice depth (up to $p \sim 0.4$) some of lattice inhomogeneities become strong enough to cause single or multiple acts of scattering (Fig. 1(a)). The physics behind this process differs substantially from well-studied case of scattering by point defects [19]. Thus, we found empirically that scattering occurs in the region with almost linear modulation of instantaneous frequency where $R(\eta) \approx \cos(\Omega_0 \eta + \beta \Omega_0 \eta^2)$, and $\beta$ characterizes the rate of the local frequency chirp. Considering interaction of incident and reflected spectral soliton components $q(\eta, \xi) = a_i(\xi) \exp(ik\eta) + a_r(\xi) \exp(-ik\eta)$ we got the system of equations for complex amplitudes $a_{i,r}$:

$$\begin{aligned} i\frac{da_i}{d\xi} &= \frac{1}{2}k^2 a_i - 2p[R_k(0)a_i + R_k(2k)a_r], \\ i\frac{da_r}{d\xi} &= \frac{1}{2}k^2 a_r - 2p[R_k(0)a_r + R_k(-2k)a_i], \end{aligned} \qquad (2)$$

where $R_k(k) = (4/\pi\beta\Omega_0)^{1/2} \cos[(\Omega_0^2 + k^2)/4\beta\Omega_0 - \pi/4]$ is the spatial spectrum of lattice with linear chirp. Strictly speaking Eq. (2) describe interaction of slant planar waves (or Fourier components of wave packet) under condition of Bragg resonance $k = \pi/\Omega(\eta)$, but it also could be applied for qualitative examination of narrow-band wave packet reflection. Analysis of Eq. (2) shows that the energy exchange distance between the incident and reflected waves is given by $L_e = \pi/[2p|R_k(2k)|]$, and that $|R_k(2k)|$ has a notable flat maximum in the vicinity $k \approx 0$ provided that $\beta = \Omega_0[\pi(1 + 4m)]^{-1}$ for $m = 0,1,2...$. The full width of this flat reflection band is given by $\Delta k = 2(2\pi\beta\Omega_0)^{1/2}$. All spectral soliton components belonging to this band will be reflected *synchronously* on $\xi$ and soliton will retain its shape upon scattering. Notice that the distance $L_e$ diminishes with growth of $p$, so that probability of soliton scattering by linearly chirped lattice fragments increases with $p$ as well. This phenomenon is entirely analogous to optical pulse reflection by linearly chirped Bragg-grating.

With further growth of the lattice depth the scenario of random soliton trapping becomes dominant (Fig. 1(b)). For the relatively deep lattices with $p \geq 1$ soliton is usually immobilized in a close proximity of launching point, while typical transverse displacement is of the order of correlation length. Soliton is typically trapped in the instantaneous lattice frequency minimum. In this region $R(\eta) \approx \cos(\Omega_0 \eta + \gamma \Omega_0 \eta^3)$, where $\gamma > 0$ characterizes the rate of quadratic FM, and the lattice spectrum is given by $R_k(k) = (3\gamma\Omega_0)^{-1/2}(\mathrm{Ai}[(3\gamma\Omega_0)^{-1/3}(\Omega_0 + k)] + \mathrm{Ai}[(3\gamma\Omega_0)^{-1/3}(\Omega_0 - k)])/2$, where Ai is the Airy-function. Considering symmetric pairs of plane waves $(a_i = a_r)$ in Eq. (2) one gets the dispersion relation $b(k) = -k^2/2 + 2p[R_k(0) + R_k(2k)]$ for propagation constant $b(k)$. It is straightforward to show that regime of normal wave diffraction is replaced by regime of anomalous diffraction for $d^2b/dk^2 \geq 0$ that occurs for broad range of spatial frequencies $k \in [-\Omega_0, \Omega_0]$ as soon as lattice depth exceeds the critical value $p > p_{\mathrm{cr}} = (8\Omega_0(3\gamma\Omega_0)^{-3/2} \mathrm{Ai}[(3\gamma\Omega_0)^{-1/3}])^{-1}$. Physically, this means that in the area of instantaneous frequency minimum the Bragg-type guiding channel is formed where soliton can be captured. The probability of soliton trapping then greatly increases as $p \to p_{\mathrm{cr}}$. Notice, that mixed localization scenarios (i.e. soliton scattering followed by its trapping) are also possible.

Figures 2(a) and 2(b) show histograms of the path-averaged soliton center displacement and its squared deviation. Such histograms provide us with the number of lattice profile realizations $N_c \leq N$ corresponding to the situation when $\eta_k^{\mathrm{av}}/\eta_0$ or $S_k^{\mathrm{av}}/S_0$ fall into fixed intervals. On the basis of Figs 2(a) and 2(b) one can draw the conclusion about key features of the probability density functions of $\eta_k^{\mathrm{av}}$ and $S_k^{\mathrm{av}}$. For instance, the histogram of $\eta_k^{\mathrm{av}}/\eta_0$ is asymmetric, its maximum corresponds to the most probable value of path-averaged soliton center displacement, and its width gives the information about the localization area. The $S_k^{\mathrm{av}}/S_0$ histogram provides information about the fraction of trajectories that differ remarkably from linear ones.

The impact of lattice depth on statistically averaged soliton center displacement $\langle\eta\rangle$ and its squared deviation $\langle S \rangle$ is illustrated in Figs. 3(a) and 3(b). Both $\langle\eta\rangle$ and $\langle S \rangle$ decrease monotonically with growth of the waveguiding parameter, but the slope of both dependencies drops off remarkably in the vicinity of $p \approx 0.25$. Careful analysis of propagation dynamics shows that it is around this value of $p$ the multiple-scattering scenario is gradually replaced with the trapping one. Therefore increase in the lattice depth leads to the qualitative modification of stochastic soliton dynamics. Notice that the variance of $\langle\eta\rangle$ also reaches its maximum value in the vicinity of the point $p \approx 0.25$ where qualitative change of propagation dynamics occurs.

Figures 3(c) and 3(d) show dependencies of $\langle\eta\rangle$ and $\langle S \rangle$ on the standard deviation of frequency fluctuations. Besides the soliton localization enhancement with growth of the disorder level, one should point out negligible localization probability at low fluctuation levels (for $\sigma \leq 10^{-2}$ at $p=1$). With increase of the disorder level beyond this critical value the localization probability growths dramatically, and random-refraction scenario is rapidly replaced with the multiple-scattering one. In the strong disorder limit ($\sigma\Omega_0 > 1$) the width of localization area saturates at the level of few correlation lengths and the center of this area is displaced in the direction of launching point. The phenomenon of the rapid growth of soliton localization probability in nonlinear optical lattices with increase of disorder level that we uncover here is reminiscent to phenomenon of Anderson localization in linear wave systems [2], and might be interpreted as smooth phase transition between *soliton- conducting* and *soliton-insulating* lattice states. Finally, to stress qualitative changes in the soliton behavior with growth of the lattice depth and disorder level, we have plotted derivatives $d_p = \eta_0^{-1}\partial\langle\eta\rangle/\partial p$ and $d_\sigma = \eta_0^{-1}\partial\langle\eta\rangle/\partial\sigma$ versus $p$ and $\sigma$ in Figs. 4(a) and 4(b), respectively. It is clear that localization process is very sensitive to small variations in waveguiding parameter in the vicinity of point $p \approx 0.1$ (Fig. 4(a)) and to variations in standard deviation of frequency fluctuations in the vicinity of $\sigma \approx 0.012$ (Fig. 4(b)).

To conclude, we have exposed the phenomenon of Anderson-type localization of walking spatial solitons in optical lattices with the random frequency modulation and encountered dramatic suppression of localization effect at low fluctuation levels. We showed that the localization process is strongly sensitive to the lattice depth since in shallow lattices moving solitons experience random refraction or/and scattering on the lattice inhomogeneities in contrast to deep lattices, where solitons are typically trapped. The estimates for parameter range where these propagation scenarios occur are given.

# Figure captions

Figure 1 (color online). Scenarios of moving soliton localization in lattices with random frequency modulation. (a) Scattering by lattice impurity at $p = 0.4$. (b) Trapping by lattice impurity at $p = 0.8$. In both cases $\Omega_0 = 8$, $\alpha_0 = 0.2$, and $L_{\text{cor}} = 2$.

Figure 2 (color online). Histograms of path-averaged soliton center displacement (a) and its squared deviation (b) at $\sigma = 0.1$, $p = 0.8$, $\Omega_0 = 8$, $\alpha_0 = 0.2$, and $L_{\text{cor}} = 2$.

Figure 3. Averaged soliton center displacement (a) and its squared deviation (b) versus waveguiding parameter at $\sigma = 0.1$. Averaged soliton center displacement (c) and its squared deviation (d) versus standard deviation of frequency fluctuations at $p = 1$. In all cases $\Omega_0 = 8$, $\alpha_0 = 0.2$, and $L_{\text{cor}} = 2$.

Figure 4. (a) Derivative $d_p$ versus waveguiding parameter at $\sigma = 0.1$. (b) Derivative $d_\sigma$ versus standard deviation of frequency fluctuations at $p = 1$. In all cases $\Omega_0 = 8$, $\alpha_0 = 0.2$, and $L_{\text{cor}} = 2$.

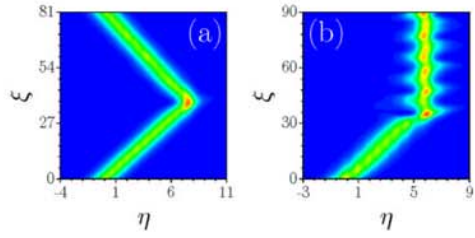

Figure 1 (color online). Scenarios of moving soliton localization in lattices with random frequency modulation. (a) Scattering by lattice impurity at $p = 0.4$. (b) Trapping by lattice impurity at $p = 0.8$. In both cases $\Omega_0 = 8$, $\alpha_0 = 0.2$, and $L_{\text{cor}} = 2$.

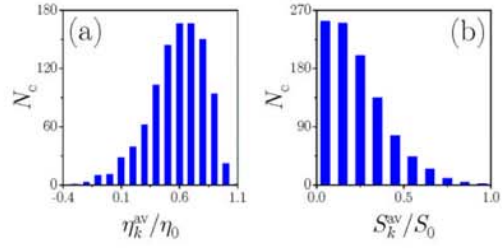

Figure 2 (color online). Histograms of path-averaged soliton center displacement (a) and its squared deviation (b) at $\sigma = 0.1$, $p = 0.8$, $\Omega_0 = 8$, $\alpha_0 = 0.2$, and $L_{\text{cor}} = 2$.

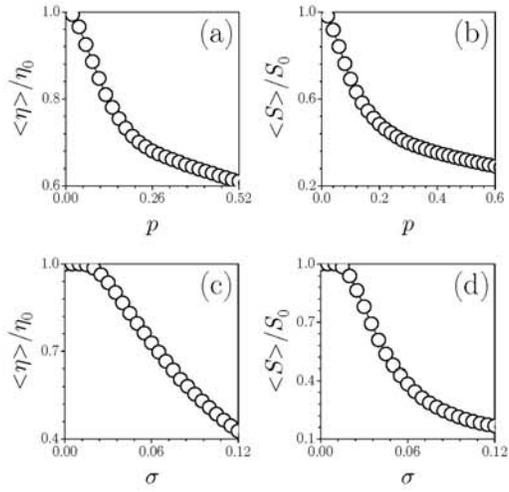

Figure 3. Averaged soliton center displacement (a) and its squared deviation (b) versus waveguiding parameter at $\sigma = 0.1$. Averaged soliton center displacement (c) and its squared deviation (d) versus standard deviation of frequency fluctuations at $p = 1$. In all cases $\Omega_0 = 8$, $\alpha_0 = 0.2$, and $L_{\text{cor}} = 2$.

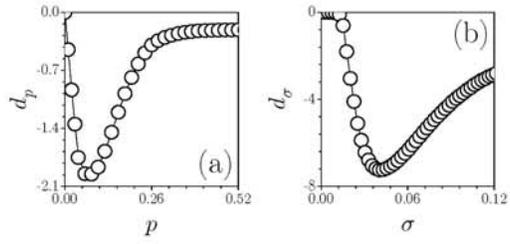

Figure 4. (a) Derivative $d_p$ versus waveguiding parameter at $\sigma = 0.1$. (b) Derivative $d_\sigma$ versus standard deviation of frequency fluctuations at $p = 1$. In all cases $\Omega_0 = 8$, $\alpha_0 = 0.2$, and $L_{\text{cor}} = 2$.